\documentclass[aps,prl,twocolumn,reprint]{revtex4}
\usepackage{graphicx}
\usepackage{epstopdf}
\usepackage{amssymb, amsmath}
\usepackage[usenames]{color}
\usepackage{bm}
\usepackage{multirow}
\usepackage{dcolumn}
\usepackage{float}
\usepackage{hyperref}
\hypersetup{
	colorlinks=true,
	linkcolor=blue,
	filecolor=gray,      
	urlcolor=blue,
	citecolor=blue,
}

\begin{document}

\title{Magnetically Controllable Topological Quantum Phase Transitions in Antiferromagnetic Topological Insulator MnBi$_2$Te$_4$}

\author{Jiaheng \surname{Li}$^{1}$}
\author{Chong \surname{Wang}$^{1,2}$}
\author{Zetao \surname{Zhang}$^{1}$}
\author{Bing-Lin \surname{Gu}$^{1,2}$}
\author{Wenhui \surname{Duan}$^{1,2}$}
\email{dwh@phys.tsinghua.edu.cn}
\author{Yong \surname{Xu}$^{1,3}$}
\email{yongxu@mail.tsinghua.edu.cn}

\affiliation{$^{1}$State Key Laboratory of Low Dimensional Quantum Physics and Department of Physics, Tsinghua University, Beijing, 100084, China\\
$^{2}$Institute for Advanced Study, Tsinghua University, Beijing 100084, China \\
$^{3}$RIKEN Center for Emergent Matter Science (CEMS), Wako, Saitama 351-0198, Japan}

\begin{abstract}
The recent discovery of antiferromagnetic (AFM) topological insulator (TI) MnBi$_2$Te$_4$ has triggered great research efforts on exploring novel magnetic topological physics. Based on first-principles calculations, we find that the manipulation of magnetic orientation and order not only significantly affects material symmetries and orbital hybridizations, but also results in variant new magnetic topological phases in MnBi$_2$Te$_4$. We thus predict a series of unusual topological quantum phase transitions that are magnetically controllable in the material, including phase transitions from AFM TI to AFM mirror topological crystalline insulator, from type-II to type-I topological Weyl semimetal, and from axion insulator to Chern insulator. The findings open new opportunities for future research and applications of magnetic topological materials.
\end{abstract}

\maketitle

The subject of antiferromagnetic (AFM) topological insulators (TIs) has attracted enormous research interests in condensed matter physics and materials science~\cite{vsmejkal2018,tokura2019}, since the first theoretical model proposed in 2010~\cite{mong2010}. AFM TIs are new states of quantum matter characterized by a $\mathbb{Z}_2$ topological invariant~\cite{mong2010,fang2013,liu2013}, in analogy to time-reversal invariant (TRI) TIs~\cite{hasan2010, qi2011,bansil2016}. Differently, their topological surface states can be inherently gapped~\cite{mong2010,fang2013,liu2013}. Remarkably, the surface gap is originated from the interplay between massless Dirac-like fermions and intrinsic magnetization, which is able to create the novel half-quantum Hall effect on the surface~\cite{qi2011,tokura2019} and the long-sought topological axion states~\cite{qi2008, wilczek1987,essin2009}. Therefore, AFM TIs provide an ideal platform to explore exotic topological quantum phenomena, including axion electrodynamics, topological magnetoelectric effects, the quantum anomalous Hall (QAH) effect and topological superconductivity~\cite{qi2011, qi2008, wilczek1987,essin2009,peng2018}. The same kind of phenomena, in principle, could be realized in TRI TIs, which, however, relies critically on extrinsic magnetic effects (such as magnetic field, doping or heterojunctions)~\cite{chang2013,nomura2011,morimoto2015,wang2015,mogi2017,mogi2017_2,xiao2018}, making corresponding material systems difficult to fabricate and control in experiments. In contrast, AFM TIs with intrinsic magnetic effects are much superior for experiments and applications. Very recently, the AFM TI states have been theoretically and experimentally discovered in van der Waals layered material MnBi$_2$Te$_4$~\cite{li2018, zhang2018, gong2018, otrokov2018}, which was proposed to host plenty of intriguing magnetic topological phases from two dimensions (2D) to 3D. The findings enable exploring the emergent quantum physics in real materials, which has stimulated intensive effort in the field~\cite{cui2019, otrokov2019, zeugner2019, peng2018, lee2018, yan2019, vidal2019, yu2019, chen2019, wu2019, zhang2019}.

One key subject essential to the study of topological physics is to effectively manipulate topological material properties and control quantum phase transitions. Various approaches have been developed by tuning electronic structure, for instance, via strain, electric gating, quantum confinement~\cite{hasan2010, qi2011,bansil2016}. For AFM TIs, there exist additional degrees of freedom in magnetic structure, which are well known to be highly tunable and play an important role in determining electronic properties~\cite{vsmejkal2018,tokura2019}. In fact, the variance in magnetic order can significantly change material symmetries. For instance, a transition from AFM to ferromagnetic (FM) state breaks $P\mathcal{T}$ symmetry (i.e. the combination of space inversion $P$ and time reversal $\mathcal{T}$), leading to spin-split bands. Moreover, a change of magnetic orientation can result in distinct symmetry selection rules and invoke different topological classifications~\cite{schnyder2008}, which could induce distinct topological phases. The influence of magnetic structure on topological material properties is thus expected to be crucial, which, however, has not been well studied for AFM TIs.

In the work, we systematically study the influence of magnetic structure on electronic and topological properties of AFM TI MnBi$_2$Te$_4$ by first-principles calculations (see Methods in Supplemental Material~\cite{SM}). We find that the interlayer coupling of MnBi$_2$Te$_4$ is strongly restricted by $P \mathcal{T}$ symmetry and can be significantly enhanced by symmetry breaking, which emphasizes the important role of $P \mathcal{T}$ symmetry in layered materials. Furthermore, we demonstrate that the change of magnetic order and orientation, which is controllable by applying magnetic field, has profound effects on orbital hybridization and band structure of MnBi$_2$Te$_4$. Noticeably, variant exotic topological states, including AFM mirror topological crystalline insulator (TCI) and type-I topological Weyl semimetal (WSM) in 3D as well as Chern insulator in 2D, can be realized in MnBi$_2$Te$_4$ by tuning magnetic structure. Unusual quantum phase transitions between these intriguing topological states are accessible, which could not only greatly enrich our understanding of magnetic topological physics but also inspire the design of new functional devices.

\begin{figure}[htbp]
	\includegraphics[width=\linewidth]{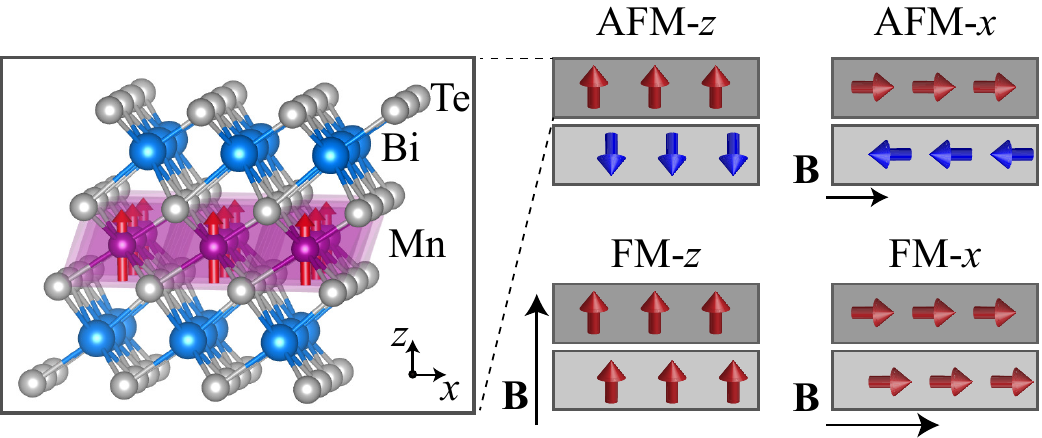}
	\caption{Variant magnetic configurations that can be realized in MnBi$_2$Te$_4$ by applying magnetic field $\bf{B}$. AFM-$z (x)$ and FM-$z (x)$ represent A-type AFM and FM order with magnetic moment oriented along the $z$ ($x$) axis, respectively. The left panel is an enlarged view of Te-Bi-Te-Mn-Te-Bi-Te septuple layer.}  \label{fig1}  
\end{figure}

\begin{table}[htbp]
	\caption{Material symmetries of MnBi$_2$Te$_4$ bulk with different magnetic configurations.}
	\begin{ruledtabular}
		{\renewcommand{\arraystretch}{1.2}
			\begin{tabular}{c c c c c c}
				\space  &      $P$     & $P\mathcal{T}$ &     $M_x$      & $C_{3z}$ & $S = T_{1/2}\mathcal{T}$	\\
				\hline
				AFM-$z$ & $\checkmark$ &    $\checkmark$      &	  $\times $    & $\checkmark$  & $\checkmark$   \\
				AFM-$x$ & $\checkmark$ &    $\checkmark$      &	  $\checkmark$ &   $\times$    & $\checkmark$   \\
				FM-$z$ & $\checkmark$  &    $\times$           &	    $\times$   & $\checkmark$  & $\times$   \\
				FM-$x$ & $\checkmark$  &    $\times$           &	  $\checkmark$ &   $\times$    & $\times$   \\
		\end{tabular}}
	\end{ruledtabular}
\label{table1}	
\end{table}

MnBi$_2$Te$_4$ is a van der Waals layered material crystallized in a rhombohedral structure with space group $R\overline{3}m$~\cite{lee2013}, formed by ABC-stacked Te-Bi-Te-Mn-Te-Bi-Te septuple layers (SLs). The ground state of this material has A-type AFM order with an out-of-plane easy axis, displaying FM order within SL and AFM order between neighboring SLs~\cite{li2018, zhang2018, gong2018, otrokov2018}, as schematically displayed in Fig. \ref{fig1}. In nonmagnetic state, there exist some key material symmetries, including rotational symmetry $C_{3z}$, mirror symmetry $M_x$, space inversion $P$ and time reversal $\mathcal{T}$~\cite{li2018, zhang2018}. Here the $x$ and $z$ axes are defined along the in-plane and out-of-plane directions, respectively (Fig. \ref{fig1}). When the AFM order is formed in MnBi$_2$Te$_4$ below the N$\mathrm{\acute{e}}$el temperature ($\sim$25 K)~\cite{gong2018, otrokov2018, cui2019, zeugner2019, lee2018, yan2019, chen2019}, $M_x$ and $\mathcal{T}$ get broken, whereas $P \mathcal{T}$ is preserved and a new symmetry $S=T_{1/2}\mathcal{T}$ emerges ($T_{1/2}$ represents a half magnetic-unit-cell translation)~\cite{mong2010}. In experiments, different magnetic orders and orientations could be realized by applying magnetic field. For instance, magnetic orientation could be tuned from out-of-plane to in-plane under a moderate magnetic field, and even further magnetic order can be driven from AFM to FM under a relatively strong magnetic field ($\sim$7 Tesla)~\cite{gong2018, cui2019, lee2018}. Different magnetic configurations (Fig. \ref{fig1}), labeled as AFM-$z$, AFM-$x$, FM-$z$ and FM-$x$ ($x$ and $z$ represent magnetic orientation), are thus accessible, which manifest different material symmetries as summarized in Table \ref{table1}. Note that ordered AFM-$x$ states are not easy to realize in MnBi$_2$Te$_4$. However, there are some intrinsic magnetic insulators in the MnBi$_2$Te$_4$ family, like XBi$_2$Te$_4$ (X=V, Ni, Eu), whose native magnetic ground state is AFM-$x$~\cite{li2018}.

\begin{figure*}[htbp]
	\includegraphics[width=\linewidth]{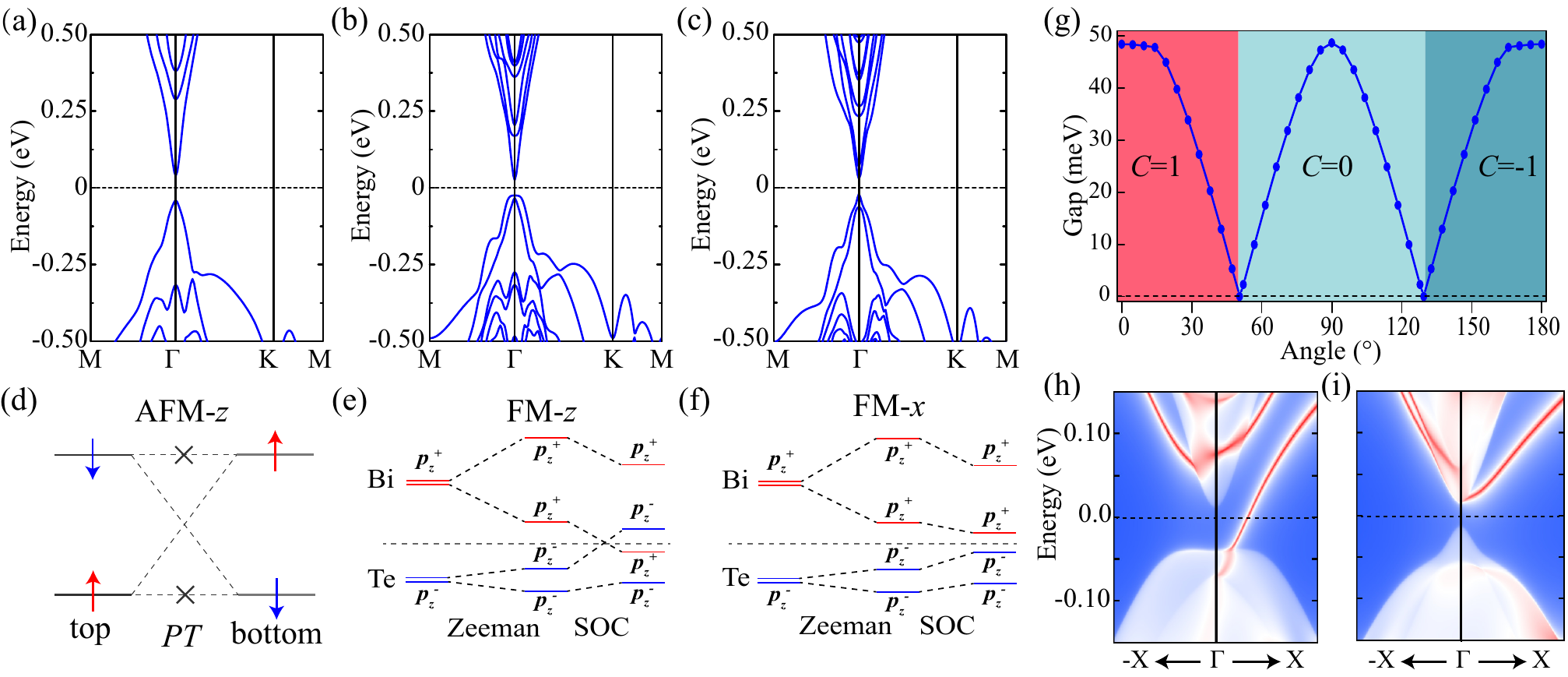}
	\caption{Properties of 2-SL MnBi$_2$Te$_4$ with different magnetic configurations. (a)-(c) Band structures for AFM-$z$, FM-$z$ and FM-$x$. (d) Schematic diagram of interlayer coupling for AFM-$z$. Coupling between energetically degenerate states, top-layer spin up (down) and bottom-layer spin down (up), is forbidden by $P\mathcal{T}$ symmetry. (e) and (f) Schematic diagrams showing the evolution of Bi-$p_z$ and Te-$p_z$ orbitals at $\Gamma$ under effective Zeeman field induced by magnetization and spin-orbit coupling for FM-$z$ and FM-$x$, respectively. (g) Evolution of band gap when varying magnetic orientation denoted by the polar angle $\theta$. Colors of shaded areas denote different topological phases of $\mathcal{C}=-1, 0, 1$. (h) and (i) Edge states for FM-$z$ and FM-$x$, respectively.} 	\label{fig2} 
\end{figure*}

Let us first consider 2-SL MnBi$_2$Te$_4$ with magnetic configurations of AFM-$z$ and FM-$z$ and discuss the influence of magnetic order on interlayer coupling. The influence is visualized by the differences in their band structures (Figs. \ref{fig2}a and \ref{fig2}b), since both systems would approach the same monolayer limit in the absence of interlayer coupling. It is well known that FM interlayer coupling gives spin-split bands, while AFM interlayer coupling introduces spin-degenerate bands~\cite{tang2016}. Generally the coupling between two separated FM SLs, each of which is spin-split by itself, would create four bonding and antibonding states. However, only two copies of states are generated by AFM interlayer coupling, which seems unusual. Detailed analysis finds that with $P \mathcal{T}$ symmetry, the top-layer spin-up state $|t,\uparrow\rangle$ and the bottom-layer spin-down state $|b,\downarrow\rangle$ are degenerate in energy, which would typically couple strongly with each other. While based on $|b,\downarrow\rangle = P \mathcal{T}|t,\uparrow\rangle$ and $(P \mathcal{T})^2 = -1$, $\langle t,\uparrow| H |b,\downarrow\rangle = \langle t,\uparrow| H P \mathcal{T}|t,\uparrow \rangle = - \langle t,\uparrow| H |b,\downarrow\rangle$, implying that the two states are forbidden to couple by $P \mathcal{T}$ (Fig. \ref{fig2}d). This unique feature, as far as we know, has not been discussed before. The analysis suggests that the interlayer coupling is strongly restricted by $P \mathcal{T}$ symmetry. This mechanism generally works for layered materials, which well demonstrates symmetry-determined interactions in quantum materials.

Conversely, interlayer coupling would be significantly enhanced by breaking $P \mathcal{T}$ symmetry, which is well demonstrated by changing magnetic order from AFM to FM. A topological band inversion between Bi $p_z^+$ and Te $p_z^-$ appears in the FM state due to the enhanced interlayer coupling (Fig. \ref{fig2}e), leading to a Chern insulator phase. Edge-state calculations find gapless chiral edge modes within the bulk gap (Fig. \ref{fig2}h), in agreement with previous results~\cite{otrokov2019}. In contrast, there is no such kind of band inversion for the AFM state, explained by restricted interlayer coupling. Therefore, when applying an out-of-plane magnetic field, a magnetic transition from AFM to FM state would be accompanied by a quantum phase transition from trivial insulator to Chern insulator.

The change of magnetic orientation also has important effects on electronic and topological properties. This is exemplified by studying 2-SL MnBi$_2$Te$_4$ with different magnetic orientations. We define a polar angle $\theta$ to quantify magnetic orientation and set $\theta = 0^\circ$ ($\theta = 90^\circ$) for FM-$z$ (FM-$x$). Their band structures are noticeably different, especially for valence bands (cf. Figs. \ref{fig2}b and \ref{fig2}c), implying strong influence of magnetic orientation on orbital hybridizations. Importantly, $M_x$ is broken for FM-$z$, while it is preserved for FM-$x$. The existence of $M_x$ requires a vanishing Hall conductance $\sigma_{xy}$~\cite{liu2013prl}, which thus suggests that FM-$x$ system has Chern number $\mathcal{C}=0$. The symmetry argument is supported by orbital analysis that finds no topological band inversion (Fig. \ref{fig2}f) as well as by edge-state calculations that show no chiral edge state (Fig. \ref{fig2}i). Remarkably, by gradually tuning magnetic orientation from out-of-plane to in-plane, a topological quantum phase transition from $\mathcal{C}=1$ to $\mathcal{C}=0$ and further to $\mathcal{C}=-1$ would occur, which is accompanied by closing and reopening of band gap (Fig. \ref{fig2}g). The magnetically tunable band gap and topological quantum phase transition could be useful for designing new functional devices, such as quantum topological transistors~\cite{molle2017}.

\begin{figure}[htbp]
	\includegraphics[width=\linewidth]{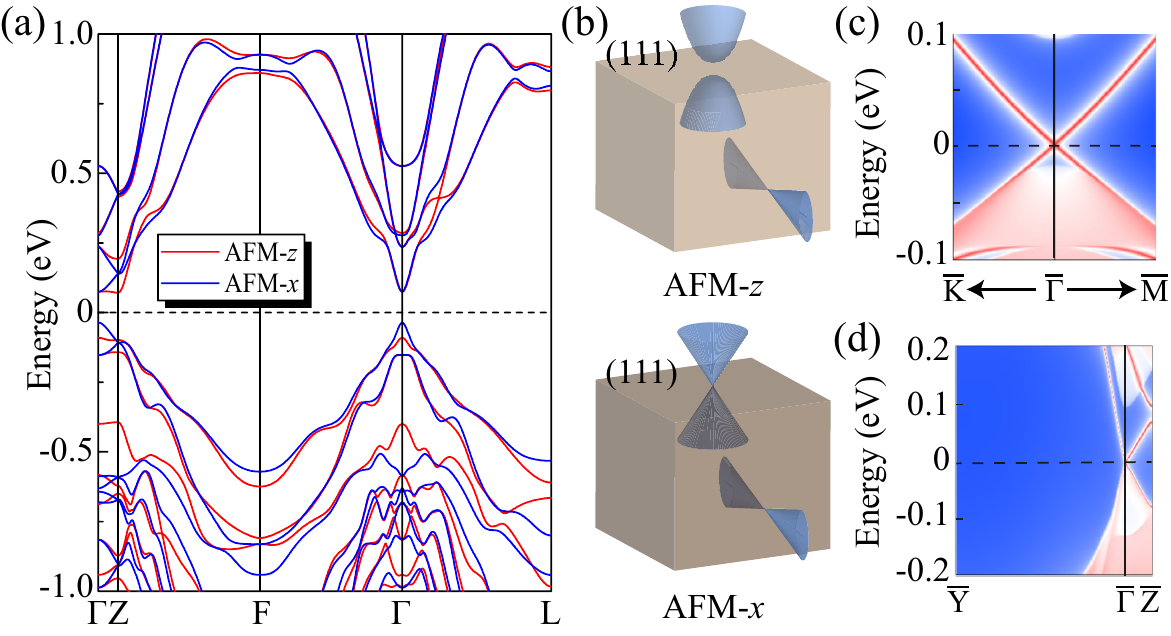}
	\caption{Electronic and topological properties of AFM MnBi$_2$Te$_4$ bulk. (a) Band structures of MnBi$_2$Te$_4$ bulk with magnetic configurations of AFM-$z$ and AFM-$x$. (b) Schematic diagrams of surface states for AFM-$z$ and AFM-$x$. (c) and (d) Topological surface states on the (111) and (110) terminations for AFM-$x$, respectively.}
	\label{fig3}   
\end{figure}

Next we turn to discuss properties of MnBi$_2$Te$_4$ bulk. Band structures of AFM-$z$ and AFM-$x$ states differ noticeably (Fig. \ref{fig3}a). This is possibly caused by their different symmetries (Table \ref{table1}), which impose different symmetry selection rules and thus result in distinct orbital hybridizations. Nevertheless, topological band inversion between Bi $p_z^+$ and Te $p_z^-$ remains unaffected. Thus, AFM-$x$ is an AFM TI characterized by $\mathbb{Z}_2 = 1$ and protected by $S$, the same as AFM-$z$. They both have gapless surface states on the side surfaces that have preserved $S$ symmetry (Fig. \ref{fig3}b). However, significant differences occur on the (111) surface. The surface states are gapped for AFM-$z$ order, which give a half-integer quantum Hall conductance $\sigma_{xy}= e^2/2h$\cite{qi2011,tokura2019}. In contrast, the surface gap of AFM-$x$, if existing, would also induces nonzero $\sigma_{xy}$, which contradicts with $M_x$ symmetry. The surface gap thus should be zero, as shown by our surface-state calculations (Fig. \ref{fig3}c). These surface states are nearly isotropic in 2D momentum space, in contrast to anisotropic Dirac-like states on the side (110) surface (Fig. \ref{fig3}d).

The AFM-$x$ state of MnBi$_2$Te$_4$ bulk hosts a new topological phase (see detailed discussions in Supplemental Material~\cite{SM}), called AFM mirror TCI, which has gapless (111) surface protected by $M_x$ symmetry and gapless side surfaces protected by $S$ symmetry. The gapless (111) surface of AFM mirror TCI is of particular interest from the point view of symmetry and topology. The surface magnetism typically introduces a mass term into the Dirac-like surface states by breaking $\mathcal{T}$. However, this mass term is enforced to zero by mirror symmetry. Based on this unique feature, one can obtain $\mathcal{T}$-breaking mass term by simply tuning mirror symmetry breaking, such as via strain, or gradual change of magnetic orientation (supplementary Fig. S1). Moreover, one may introduce another type of mass term, for instance, by superconducting proximity effect, which could induce topological superconductivity~\cite{peng2018}. Therefore, rich topological quantum physics would emerge by controlling symmetry breaking. Note that some intrinsic magnetic insulators, like XBi$_2$Te$_4$ (X=V, Ni, Eu), have magnetic ground state AFM-$x$~\cite{li2018}. The novel AFM mirror TCI phase might be realized intrinsically in these candidate materials.

\begin{figure}[htbp]
	\includegraphics[width=\linewidth]{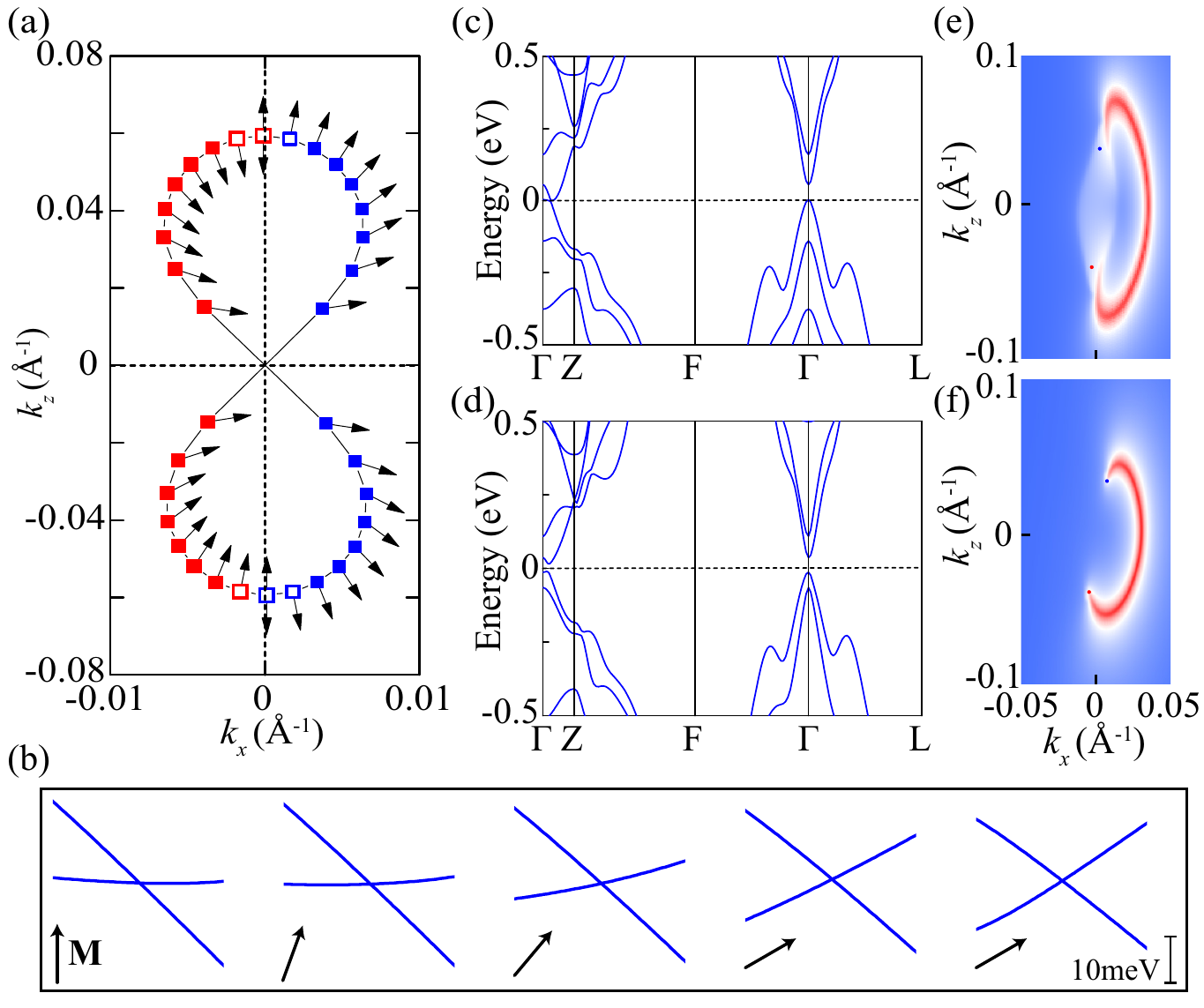}
	\caption{Tunable Weyl points (WPs) in FM MnBi$_2$Te$_4$ bulk. (a) Evolution of WPs in momentum space when magnetic orientation (denoted by black arrows) rotates from out-of-plane to in-plane direction in the $k_x$-$k_z$ plane. The closed and open squares denote type-I and type-II WPs, respectively. The blue and red colors represent topological charges of +1 and $-1$ carried by WPs. (b) The zoom-in calculated band structures around WPs along the $k_z$ direction for different magnetic orientations (denoted by black arrows) with polar angles equal to 0$^\circ$, 20$^\circ$, 40$^\circ$, 60$^\circ$ and 80$^\circ$, respectively. (c,d) Band structures and (e,f) surface states of FM bulk MnBi$_2$Te$_4$ with magnetic orientation angles of 10$^\circ$ (upper) and 50$^\circ$ (lower), which correspond to type-II and type-I WSMs, respectively.}	
	\label{fig4} 
\end{figure}

Topological properties of bulk MnBi$_2$Te$_4$ greatly change when tuning magnetic order from AFM to FM. The FM-$z$ state corresponds to type-II WSM~\cite{li2018,zhang2018}, in which Lorentz invariance is violated and some parts of electron pockets are located below hole pockets in the Weyl cone~\cite{soluyanov2015}. Importantly, only one single pair of Weyl points (WPs) exist in the material, which are located along the $\Gamma$-$Z$ direction as ensured by $C_{3z}$ rotational symmetry and related to each other by $P$~\cite{li2018,zhang2018}. When rotating magnetic orientation away from $z$ axis by applying magnetic field, $C_{3z}$ gets broken. Hence, WPs would shift away from $\Gamma$-$Z$ to general $\bf{k}$ points. Figure \ref{fig4}a displays the motion of WPs in momentum space when varying magnetic orientation (see detailed data in Supplemental Material~\cite{SM}). Quantum phase transition from type-II to type-I WSM phase happens at $10^\circ<\theta<20^\circ$. When $\theta = 90^\circ$ (FM-$x$), two WPs eventually merge, leading to annihilation of Weyl fermions and a topological transition to trivial phase. Such kind of phase transition behaviors are rationalized by the fact that Zeeman field induced band splitting is much larger for FM-$z$ than for FM-$x$ in this layered structure, and thus varying $\theta$ considerably changes band dispersions near WPs (Fig. \ref{fig4}b).

Remarkably, the type-I WSM phase in MnBi$_2$Te$_4$ is the simplest WSM, which has only a single pair of WPs located exactly at the charge neutral point. For comparison, we show band structures for $\theta = 10^\circ$ and $\theta = 50^\circ$ as representative examples of type-II and type-I WSMs (Figs. \ref{fig4}c and \ref{fig4}d), and present their calculated surface states on the ($1\overline{1}0$) termination (Figs. \ref{fig4}e and \ref{fig4}f). While Fermi arcs and  residual electron pockets around $\Gamma$ simultaneously appear at the isoenergy level of WPs in the type-II case, only Fermi arcs with no interfering signal from bulk states can be obtained in the type-I case. In contrast to other WSM materials with multiple pairs of WPs and complex Fermi surfaces~\cite{wan2011, weng2015, soluyanov2015, armitage2018}, this material hosts the simplest WSM phase, which is more advantageous for transport measurements and other studies of Weyl physics.

\begin{figure}[htbp]
	\includegraphics[width=1.0\linewidth]{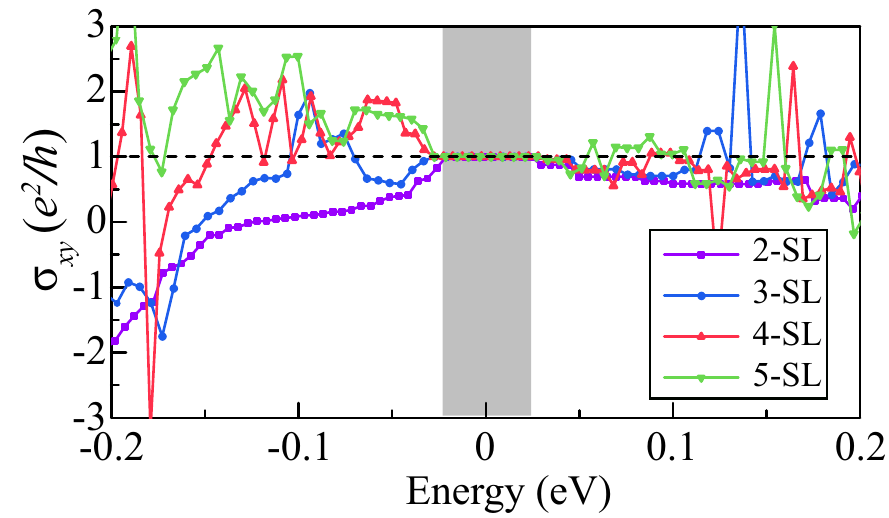}
	\caption{Anomalous Hall conductance of few-layer FM MnBi$_2$Te$_4$ as a function of Fermi energy.}
	\label{fig5} 
\end{figure}

Finally we discuss topological properties of MnBi$_2$Te$_4$(111) films in the FM-$z$ state (see results for other magnetic states in Supplemental Material~\cite{SM}). Their ultrathin films show similar band structures (supplementary Fig. S3), which are in the Chern insulator phase as shown by our calculations of anomalous Hall conductance (Fig. \ref{fig5}) and bandgap as a function of spin-orbit coupling strength (supplementary Fig. S4). Chern numbers and bandgaps (about 50-80 meV) for varying thickness in the FM-$z$ state are summarized in supplementary Table S3. In principle, thick films can have high Chern numbers, which is an inherent feature of topological WSM. In the long wavelength limit, Chern number of MnBi$_2$Te$_4$ films is estimated to be $\mathcal{C}\approx |k^z_{\textrm{W}}| d/\pi$, where $k^z_{\textrm{W}}$ is the $z$-component wave vector of WP in MnBi$_2$Te$_4$ bulk and $d$ is the film thickness. Therefore, when changing magnetic order from AFM to FM, MnBi$_2$Te$_4$ films can display topological quantum phase transition from axion insulators to Chern insulators of varying Chern numbers.

In conclusion, through an example study of MnBi$_2$Te$_4$, we demonstrate that the symmetry, orbital hybridization and band dispersion of AFM TI are highly controllable by magnetic field, thus leading to quantum phase transitions to variant exotic topological phases, such as AFM mirror TCI and type-I topological WSM. This creates new opportunities for future research. For instance, the tunable Weyl fermions in FM MnBi$_2$Te$_4$ can host some novel quantum phenomena, like the black-hole-horizon analogy emerging at the Lifshitz transition between type-I and type-II WSM phases~\cite{volovik2017,huang2018_2} and 3D quantum Hall effects~\cite{zhang2019nature,tang2019} without Landau levels.

\emph{Note added:} Recently, magnetic field induced Chern insulators ~\cite{deng2019} and quantum phase transition from axion insulator to Chern insulator ~\cite{liu2019} were realized experimentally in few-layer MnBi$_2$Te$_4$.

We thank Yayu Wang and Jinsong Zhang for helpful discussions. This work is supported by the Basic Science Center Project of NSFC (Grant No. 51788104), the Ministry of Science and Technology of China (Grants No. 2016YFA0301001, No. 2018YFA0307100, and No. 2018YFA0305603), the National Natural Science Foundation of China (Grants No. 11674188 and No. 11874035) and the Beijing Advanced Innovation Center for Future Chip (ICFC).


\end{document}